\documentstyle[12pt,aasms4]{article}

\newcommand{\degrees}{\mbox{$^{\circ}$}}    
\newcommand{\NH}{\mbox{${\rm N_H}$ }}       
\newcommand{\NHunits}{\mbox{$ 10^{20}~{\rm cm}^{-2}$}}

\begin{document}

\title {AN ULTRA-SOFT X-RAY SOURCE (RX~215319-1514) AND DISCOVERY OF A NARROW LINE SEYFERT 1 GALAXY}
\author {Kulinder Pal Singh \altaffilmark{1,3} \& Laurence R. Jones \altaffilmark{2,4}}

\altaffiltext{1}{Department of Astronomy \& Astrophysics, Tata Institute of Fundamental Research, Mumbai 400005, India}
\altaffiltext{2}{Department of Physics \& Astronomy, University of
Birmingham, Birmingham B15 2TT, U.K.}
\altaffiltext{3}{singh@tifr.res.in}
\altaffiltext{4}{lrj@star.sr.bham.ac.uk}

\slugcomment{Version Date: September 20, 1999}

\begin{abstract}
We present results on the identification of the optical counterpart 
of an ultra-soft X-ray source discovered with $ROSAT$.
Two optical candidates -- a galaxy and a star, are found within 
the error circle of the X-ray source position.  
Optical spectroscopy of the two candidates reveals that a) the galaxy
is a narrow-line Seyfert type 1 galaxy, and b) the star
is a late A-type or an early F-type.  The F$_x$/F$_{opt}$ ratio 
is too high for the star to be the counterpart of 
the X-ray source, but consistent with that for an active galaxy.
Although higher resolution X-ray imaging of the region is needed
to definitely settle the question of the counterpart of the X-ray source,
the narrow-line Seyfert 1 galaxy is the best candidate.  
The spectral properties of the newly discovered narrow-line Seyfert 1  
galaxy are also presented, including its
extreme X-ray power law spectral index of $\Gamma$ $\geq$4.
\end{abstract}


\section{INTRODUCTION}
The catalogue of ultra-soft X-ray sources of Singh et al. (1995)
based on WGACAT (White, Giommi, \& Angelini 1994),
consists mostly of white dwarfs, cataclysmic variables and some stars.
A few extragalactic sources (galaxies and QSOs) which have very steep
X-ray spectra and lie in the direction of low interstellar hydrogen
column density are also included.  These few sources are quite unusual
since the ratio of soft (0.1--0.4 keV) and medium energy (0.4--0.9 keV)
fluxes is many times larger than for a typical active galaxy or 
an AGN that they resemble.
Although the soft excesses over and above the canonical power-law
in Seyferts  can cause these galaxies to appear very soft, recent 
observations with $ROSAT$ have shown the existence of 
intrinsically steep spectrum Seyferts (Brandt, Pounds, \& Fink 1995;
Boller, Brandt, \& Fink 1996; Laor et al. 1997) 
which have very narrow emission lines and are, therefore, classified
as narrow-line Seyfert 1 galaxies.  As part of our programme to 
optically identify
the counterparts of ultra-soft sources in the catalogue of 
Singh et al. (1995), we have discovered a narrow-line Seyfert 1 galaxy 
which is a strong candidate for being a counterpart of an ultra-soft 
X-ray source viz., WGA~J2153.3-1514 (RXJ~2153.3-1514).

In this paper, we present our analysis of the archival X-ray data 
on the  ultra-soft X-ray source and optical spectroscopic
observations of two bright objects within the error circle 
of the ultra-soft X-ray source.  One of these objects is found to be 
a bright narrow-line Seyfert 1 galaxy at a redshift of 0.0778, and the 
other is a star. 

\section {X-RAY DATA AND ANALYSIS}

The region of the sky containing this source was observed 
with the $ROSAT$ (Truemper et al. 1983) PSPC detector twice, 
once from 1992 November 19 to 29 and then from 1993 May 6 to 11. 
ROSAT X-ray data corresponding to these observations were obtained 
from the public archives maintained at the High Energy Astrophysics Science
Archive Research Center (HEASARC) in USA, and analysed by us.  
The data are referred by the sequence numbers rp800227n00 and rp800227a01
(rp800227p and rp800227p-1 in the public archives in Germany)   
for the 1992 and 1993 observations respectively.  
The  effective exposure times were 9040~s and 17436~s
in the first and second observation respectively.
Abell 2382 was the target source in both the observations.   
An unidentified ultra-soft X-ray source was detected $\sim$ 41$\arcmin$ 
from the centre of the field in the two sets of observations. 
The source was not obstructed by the PSPC window support structure.
The source is, however,  rather weak with a total of 203$\pm$22 counts 
in the first observation, and 375$\pm$31 counts in the second
observation in the $ROSAT$ energy band of 0.1 - 2.4 keV. 
The source counts were obtained from a circle of radius 2.7$\arcmin$
centered on the peak position, and after subtracting the background 
estimated from an annular region with the same center and
with inner and outer radii of 4.2$\arcmin$ and 8.4$\arcmin$
respectively. This was done using the $imcnts$ program in 
the PROS software package.  The large radius for source extraction
was necessitated due to the broadening  of the point spread function (psf)
at  large off-axis angle from the centre of the field of view of the
telescope.  The optimal extraction radius was selected by 
changing the radius and finding the maximum in the total source counts.
The background estimate was checked by increasing the inner and outer
radii by a few arcmins and no significant change was found.
The X-ray intensity in the two observations is, thus, almost unchanged.

We have extracted data from identical sections of the X-ray image 
around the ultra-soft source in the two observations.
Due to the large off-axis angle of the X-ray source, 
the psf of the telescope is as large as 1.7$\arcmin$ (half energy radius) 
leading to a considerable uncertainty in the position of the source 
in both the observations.  
In addition, the weakness of the source does not allow us to correct 
for any residual pointing errors in the two observations,
smaller than the psf.
X-ray contour maps made from the individual observations
of the source are shown in Figures 1 and 2 after overlaying on an 
optical image of the source that was extracted from the Digital Sky 
Survey (DSS).
The raw image was smoothed with a Gaussian 
($\sigma$= 20 pixels = 10$\arcsec$) before plotting in both the cases.  
The structure seen in the contour maps is unlikely to be real
as these can arise due to the large off-axis angle of the source
(see Hasinger et al. 1993).
A galaxy (V=14.7 mag) and a star (V=11.72 mag) can both be seen close 
to the central maximum in the X-ray maps in Figs. 1 and 2.  
The position of the galaxy, as
measured from the DSS, is RA(2000) = 21$^h$ 53$^m$ 19$^s$ and
Declination (2000) = -15\degrees 14$\arcmin$ 20$\arcsec$.  The star is
identified in the HST Guide Star Catalog as GSC~0637600659 and its
position is given as RA(2000) = 21$^h$ 53$^m$ 16$^s$.8, Declination(2000) =
-15\degrees 14$\arcmin$ 22.9$\arcsec$.   
The X-ray maxima in the two images do not coincide.
The separation between the two maxima is, however,  only 0.6$\arcmin$.
It is quite clear from Fig. 1 that the galaxy is an X-ray source.
X-ray emission from the star, although somewhat unlikely (see \S 4 below), 
cannot be ruled out, however.  It is very difficult to quantify the 
contribution of star to the observed X-ray emission, however, using 
a radial profile of the source we estimate that $\leq$40\% of the 
flux may come from the star if it is indeed an X-ray emitter.
No radio source has been reported within 5$\arcmin$
radius from the X-ray source.

A photon energy spectrum was accumulated from the 
on-source counts obtained from a circular region on the sky, 
having a radius of 2\farcm7, to account for the large point spread function.
The corresponding background spectrum was accumulated from several 
neighbouring regions at nearly the same off-axis angle as the source.  
This procedure was applied to the data from both the observations separately.
The X-ray spectra from the two observations are nearly identical and
are shown together in the upper panel of Figure 3, after regrouping
of channels to improve the statistics.
Almost all the counts in the spectrum are below 0.5 keV.

We used the XSPEC (Version 10.0) spectral analysis package to
fit the data with spectral models.   This requires a knowledge
of the response of the telescope and the detector. 
Using the source spectral files and information about the source
off-axis angle contained within them, we generated an auxiliary response file of the
effective area of the telescope, by using the program $pcarf$ in the
FTOOLS (version 4.0) software package.  This program uses the
available off-axis calibration of the telescope, and the amount of
scattering of X-ray photons and their dependence on energy are taken
into account.  The response matrix $pspcb-gain2-256.rmf$, as provided
by the ROSAT GOF at HEASARC, was used to define the energy response of 
the PSPC.
The $ROSAT$ PSPC pulse height data were re-grouped to have a
minimum of 25 counts per grouped channel and then used for fitting
spectral models so that the $\chi^2$ test for finding best fit models
could be applied meaningfully.  Channels with no useful data, after
background subtraction, were ignored.  The spectra from the
two observations were fitted separately as well as jointly.
The two spectra were kept separate to avoid the introduction of 
any systematic errors.  The results of a joint analysis are as follows:

Using a power-law model and absorption due to an intervening medium 
using absorption cross-sections as given by Morrison \& McCammon (1983), 
we obtain a best fit ($\chi^2_{min}$= 24.05 for 33 degrees of freedom)
for a photon index ($\Gamma$) of 6.4$^{+5}_{-1.5}$ and  
\NH=4$^{+3.9}_{-1.4}\times$\NHunits.  The errors quoted, here and
below, are with 90\% confidence based on $\chi^2_{\rm min }$+2.71.
The best-fit value for \NH is about the same as the 21-cm value 
(\NH=4$\times$\NHunits) from radio observations in this direction, 
indicating that all the X-ray absorption is due to matter in our 
own Galaxy.  Based on the best fit model parameters, the total X-ray flux 
from the source is estimated to be 9.5$\times$10$^{-14}$ ergs 
cm$^{-2}$ s$^{-1}$ with interstellar absorption, 
and 1.0$\times$10$^{-11}$ ergs cm$^{-2}$ s$^{-1}$ 
without intervening absorption in the 0.1--2.0 keV energy band.  
The large difference between the absorbed and unabsorbed
fluxes is due to the very steep spectral index.
The best fit power-law model is shown as a histogram in Fig. 3.
We have derived the 68\%, 90\%, and 99\% confidence regions for 
$\Gamma$ and \NH, based on the joint analysis of the two spectra,
as well as from the analysis of individual spectra, and 
these confidence regions are shown as contour diagrams in Figure 4.
This Figure shows that the results from these analyses are consistent with 
each other.

Using black-body or thermal plasma emission models gives equally good
fits to the spectra.   It is, however, difficult to constrain both the
temperature and \NH independently.  Keeping \NH fixed at the
21-cm value, we get the best fit ($\chi^2_{min}$= 22.7 for 34 degrees of 
freedom) value of kT = 30$^{+9}_{-9}$ eV, and 0.1--2.0 keV X-ray flux
of 7.5$\times$10$^{-14}$ ergs cm$^{-2}$ s$^{-1}$ (with absorption)
and 2.7$\times$10$^{-12}$ ergs cm$^{-2}$ s$^{-1}$ (without absorption). 

Using a ``mekal'' optically thin thermal plasma model (Kaastra 1992) 
with solar abundance also gives an equally good fit to the data
with $\chi^2_{min}$= 24.6 for 34 degrees of freedom.  The best fit
kT = 54$^{+11}_{-12}$ eV for fixed \NH=4$\times$\NHunits.  
The 0.1--2.0 keV X-ray flux for these parameters is 1.4$\times$10$^{-13}$ 
ergs cm$^{-2}$ s$^{-1}$ (with absorption) and 
5.3$\times$10$^{-12}$ ergs cm$^{-2}$ s$^{-1}$ (without absorption).
As can be seen from the above analyses the flux values are very 
sensitive to the assumed value of \NH and the shape of the spectrum.

\section {OPTICAL OBSERVATIONS AND ANALYSIS}

Low resolution optical spectra of candidate counterparts were obtained
at the Lick 3m telescope on 1995 August 2. A long slit was 
positioned across the 11th mag star and the nucleus of the 14th mag galaxy, 
near the peak of the X-ray emission in Fig 1. The slit 
position angle was 71.1 degrees. The Kast spectrograph
was used with a dichroic beam splitter which operated at 5500 \AA,
sending the blue and red light to blue and red  spectrograph arms,
each with Reticon CCD detectors.
A slit width of 2.5 arcsec and a grism (blue) and a grating (red)
both of 600 lines/mm gave a resolution of 6.5 \AA~(FWHM). 
The exposure time was 400 sec and the airmass was 2.7.

Flat-fielding and wavelength calibration were performed using 
appropriate quartz lamp and arc lamp exposures, which were 
taken frequently through the night.
The quartz lamp exposures were scaled and 
combined to reject cosmic rays and to improve the signal-to-noise
ratio. The wavelength calibration was accurate to $\sim$2\AA~, as 
measured from night sky lines. 
Approximate flux calibration was performed using standard $IRAF$ 
procedures via 
observations of two Kitt Peak spectrophotometric standard stars.
We were not able to correct for the loss of blue light due to 
atmospheric refraction, so the flux calibration is very approximate.
Some residual continuum fluctuations are visible at $\lambda>5500$ \AA~ in
the spectrum of the star. 
Atmospheric absorption bands at $\lambda>$6800 \AA~were removed in the
galaxy spectrum using
high signal-to-noise observations of bright stars observed at a similar 
airmass as the target galaxy.

The spectrum of the galaxy is shown in Figure 5. The gap at 
5500 \AA~is due to the effect of the dichroic. 
Although the spectrum is noisy, [O III] (5007\AA), H$\alpha$ and 
[N II] emission lines are visible. 
The redshift derived from Gaussian fits to these three lines is 
0.0778$\pm$0.0002. The H$\alpha$ + [N II] region was fit with three
emission lines with fixed centres but variable widths and 
intensities. The resulting line parameters are given in Table 1.
Errors on the line widths were determined from 100 Monte-Carlo 
simulations of the region.

The parameters of the [N II] 6549 line (and the [O III] line) are not 
well constrained (the [N II] 6549 FWHM is 
smaller than the instrumental resolution). The H$\alpha$  line, however,
is clearly resolved. After correcting for the  instrumental resolution,
the intrinsic FWHM is 880$\pm$50 km s$^{-1}$. 
This value, together with the very soft X-ray spectrum,  is indicative of
a narrow-line Seyfert 1 galaxy (Boller, Brandt \& Fink  1996). 
We note that although H$\beta$, rather than H$\alpha$, line widths are
usually quoted for NLS1s (because
H$\beta$ remains unblended at lower resolution), 
Osterbrock \& Pogge (1985) found that the H$\alpha$ and
H$\beta$ NLS1 line widths were very similar. 

In order to confirm the NLS1 classification, the spectrum was smoothed
with a wide boxcar function of width 50$\AA$ to increase the 
signal-to-noise in broad spectral features, such as the blends of FeII 
emission lines which form part of the  
NLS1 definition of e.g., Goodrich (1989). The resulting
spectrum is shown in Figure 6.  Redshifted FeII 4570$\AA$, with the 
expected width of $\approx$200$\AA$, is clearly visible, along with 
H$\beta$, [O III] and a marginal detection of the FeII 5190-5300$\AA$ blend.
The presence of the FeII features confirm the classification as
a NLS1 galaxy. The continuum is, however, too poorly defined to reliably
measure the emission line parameters at the blue end of the spectrum.
A higher signal-to-noise spectrum is clearly needed to investigate the 
line parameters in more detail.

The star near the X-ray source position, known as GSC~0637600659, has
optical magnitudes given as Q$_{V}$=11.72 and  Q$_{B}$=11.82, in
the HST GSC catalog.  The spectrum of the star is shown in Figure 7.  
No emission lines characteristic of an accreting cataclysmic 
variable are seen in the spectrum. There is also no evidence
for chromospheric activity.
Based on the spectral characteristics the star 
is most probably of late-A to early-F spectral type.

\section {DISCUSSION}

The position of the ultra-soft X-ray source is not well determined
in the present observations making the task of optical identification
difficult, although two bright optical objects are seen near the X-ray peak
position.   Multiple sources being responsible for the X-ray emission are,
however, unlikely since the log N -- log S relation (Hasinger, Schmidt, \&
Truemper 1991) predicts only 10$^{-1}$ sources per square degree
with flux (S) $\geq$3$\times$10$^{-13}$ ergs cm$^{-2}$ s$^{-1}$
(mean separation $\approx$3\degrees), and
10$^{-3}$ sources per square degree with flux 
$\geq$3$\times$10$^{-12}$ ergs cm$^{-2}$ s$^{-1}$ (mean separation
$\approx$30\degrees).  The probability
of two X-ray sources, each with a flux similar to
the observed flux, having a separation of
5$\arcmin$--10$\arcmin$ is, therefore, quite low.

The ratio of X-ray flux (F$_x$) to optical flux (F$_v$) is a very useful
tool in the process of optical identification of X-ray sources
(see Maccacaro et al. 1988).  Using the energy range of 0.3--3.5 keV
to calculate the X-ray flux, Maccacaro et al. give the following
relationship to estimate the ratio log(F$_x$/F$_v$):

\centerline{log(F$_x$/F$_v$) = log~F$_x$ + m$_v$/2.5 + 5.37}

Based on allowed spectral models fitted to the X-ray spectrum,
the log~F$_x$ in the 0.3--3.5 keV energy range is estimated to be
-13.32 (blackbody spectrum), -13.07 (power-law), and -12.82 (thermal plasma).
(The 0.1--2.0 keV fluxes quoted earlier are higher because of the 
steepness of the spectrum and sensitivity to \NH).
Thus, we find that if the X-ray emission is attributed to the star then 
the flux ratio, log(F$_x$/F$_v$), is in the range  -3.26 to -2.76.  
Comparing these values with the nomograph given in Figure 1
of Maccacaro et al. (1988), we can see that the flux values and the ratios
indicate the star should be of G, K or M-type.  Other possibilities
are active stars in  RSCVn-type binaries (late G to K subgiants or giants)
or among rapid rotators (usually dwarf K or dwarf M-types).
The optical spectrum (Fig. 7) taken by us, however, clearly shows 
the star to be a normal A-F type.
Besides, active stars do not usually have such a soft X-ray spectrum
(Singh et al. 1999).  
Even solar-types have kT=0.3 keV and L$_x$/L$_{v}$ of $\sim$ 
10$^{-7}$--10$^{-6}$ (Mewe et al. 1998).
For the star to be as soft as it is in X-rays it must harbor either 
a white dwarf or be a cataclysmic variable of some kind,
but the optical spectrum shows only the signature of a normal star.

The narrow-line Seyfert 1 (NLS1) galaxy  has V $\simeq$ 14.7 
and if it is the source of X-rays then log (F$_x$/F$_v$) is in the 
range -2.07 to -1.57, if the total flux can be attributed to it.  
This ratio is well in the range observed for galaxies and 
low luminosity AGN (even if only 60\% of the total flux is from
the galaxy).  The optical spectrum of the galaxy shows 
an H$\alpha$ emission line with a width intermediate between those
of Sy1 and Sy2 galaxies, typical of NLS1s. Such galaxies are known to have 
steep soft X-ray spectra.   
For the observed redshift and assuming H$_o$=75 km s$^{-1}$ Mpc$^{-1}$,
q$_o$=0 in a Friedmann cosmology, the luminosity distance
of the Seyfert galaxy is 323.3 Mpc.  The (unabsorbed) 
X-ray luminosity of the Seyfert galaxy in the 0.1--2.0 keV energy 
band is  3.4$\times$10$^{43}$ ergs s$^{-1}$ for a blackbody spectrum,
1.25$\times$10$^{44}$ ergs s$^{-1}$ for a power-law type spectrum, and
6.6$\times$10$^{43}$ ergs s$^{-1}$ for a thermal plasma.
This X-ray luminosity is quite typical of Seyfert galaxies.
The observed X-ray source is, therefore, very likely to be associated 
with this narrow emission line galaxy at a redshift of 0.0778.

The value of the X-ray spectral index,  $\Gamma$ $\geq$ 4 (99\% confidence),
is extreme even for the NLS1 class of AGN, being steeper than any 
of the galaxies in the compilation of Boller et al. (1996).
The position on the $\Gamma$-H$\beta$ line width plane 
(Fig. 8 of Boller et al) is, however, consistent with an extrapolation 
based on the other NLS1 galaxies. 

\section {CONCLUSIONS}

A narrow-line Seyfert 1 galaxy has been discovered in the error circle
of an ultra-soft X-ray source.  Although higher spatial resolution
and broad-band X-ray observations are needed to be certain about the
association of this galaxy with the ultra-soft X-ray source, based on
the spectral characteristics it appears to be the better candidate 
for being the counterpart of the X-ray source than the bright
A--F type star which is also in the error circle.  Assuming this
association to be true, the 0.1--2.0 keV X-ray luminosity of the 
NLS1 is in the range of (3.4--12.5)$\times$10$^{43}$ ergs s$^{-1}$, 
and its soft X-ray spectral index is extremely steep ($\Gamma$ $\geq$ 4).

\acknowledgments
This research has made use of the PROS software package provided
by the ROSAT  Science Data Center at Smithsonian Astrophysical
Observatory, and  the FTOOLS software package provided  
by the High Energy Astrophysics Science Archive Research Center 
(HEASARC) of NASA's Goddard Space Flight Center.

\clearpage
\setcounter{page}{14}
\begin{figure}[h]

\figurenum{1}
\caption
{X-ray image of RX~215319-1514.6 as observed with the $ROSAT$ PSPC 
in 1992 observations after smoothing
with a Gaussian ($\sigma$=10\arcsec) and overlaid
on the optical image (gray) obtained from the Digital Sky Survey.   
X-ray contours are plotted at 30, 40, 50, 70, 80, 90 and 95\% of the peak 
brightness of 0.0049 counts pixel$^{-1}$.
(1 pixel =0.5\arcsec$\times$0.5\arcsec)}

\figurenum{2}
\caption
{X-ray image of RX~215319-1514.6 as observed with the $ROSAT$ PSPC 
in 1993 observations after smoothing
with a Gaussian ($\sigma$=10\arcsec) and overlaid
on the optical image (gray) obtained from the Digital Sky Survey.   
X-ray contours are plotted at 40, 50, 70, 80, 90 and 95\% of the peak 
brightness of 0.0066 counts pixel$^{-1}$.
(1 pixel =0.5\arcsec$\times$0.5\arcsec)}

\figurenum{3}
\caption
{$ROSAT$ PSPC X-ray spectra of the ultra-soft X-ray source obtained
from two observations and fitted jointly with a power-law model are
shown in the upper panel. 
The best fit power-law model from a joint fit is shown as histograms.
The significance of the residuals between the observed data points and 
the model are shown in the lower panel.}

\figurenum{4}
\caption
{Allowed ranges of the power-law photon index  $\Gamma$ and N$_H$
for 68\%, 90\%, and 99\% confidence based on counting statistics.
The `+' marks the best fit value.}

\figurenum{5}
\caption
{Optical spectrum of the active galaxy. Prominent emission lines 
are labelled. The gap at 5500\AA~ is instrumental. The spectrum has
been smoothed with a box of size 3 pixels (6.9\AA) corresponding to
the instrumental resolution.  }

\figurenum{6}
\caption
{Same as in Fig.5 but smoothed with a box of size 50\AA.}

\figurenum{7}
\caption
{Optical spectrum of the star, GSC~0637600659.}

\end{figure}

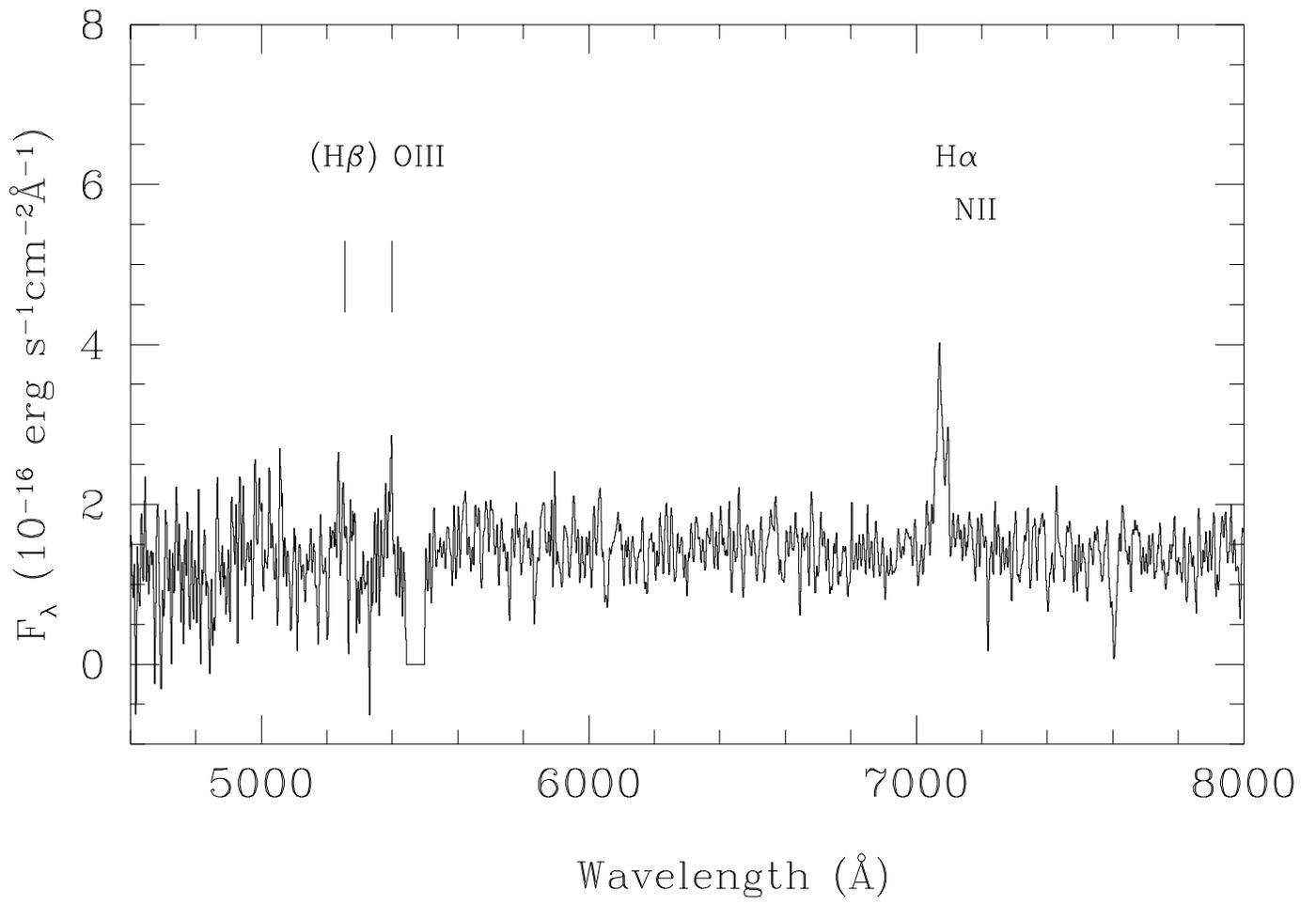
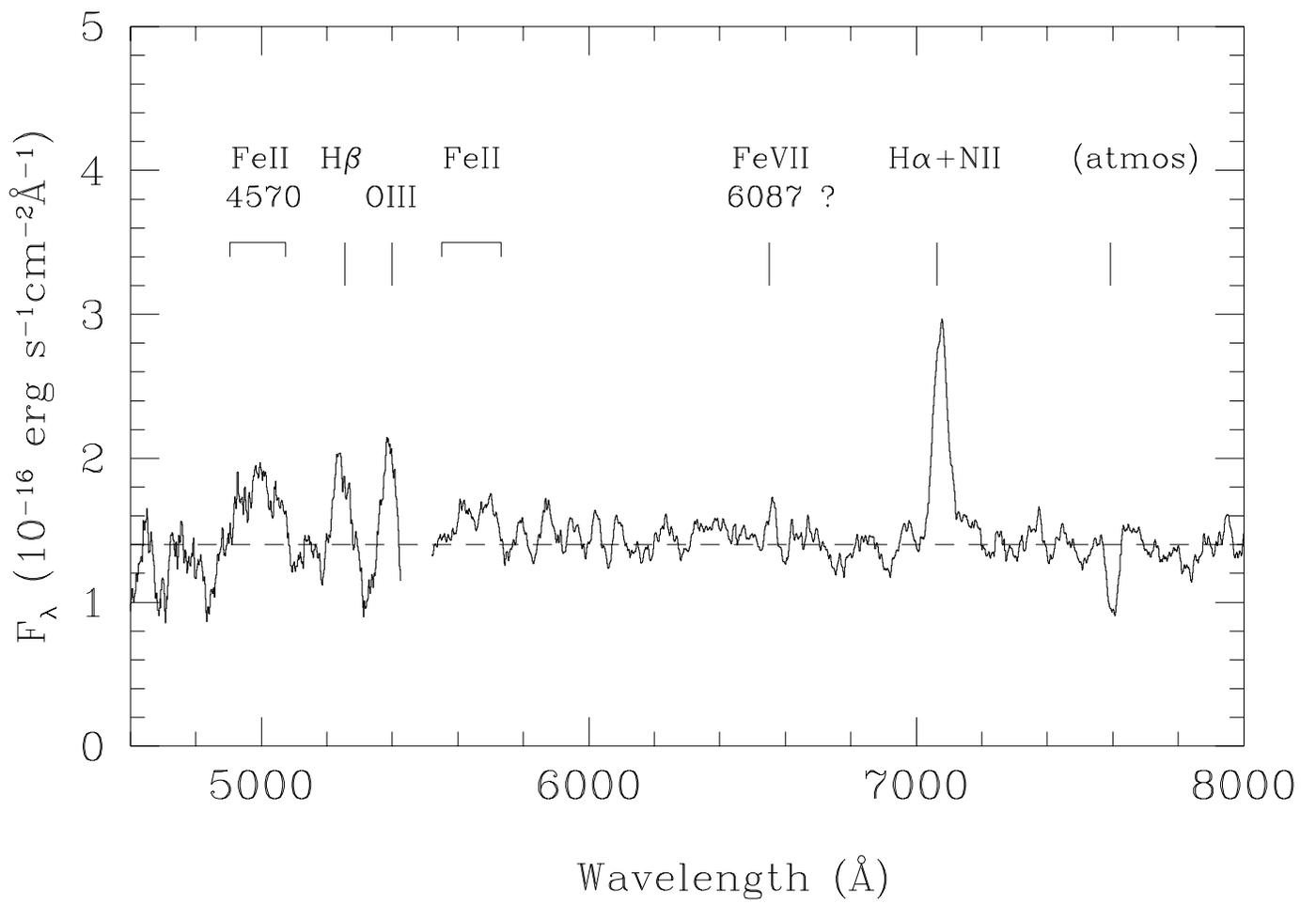
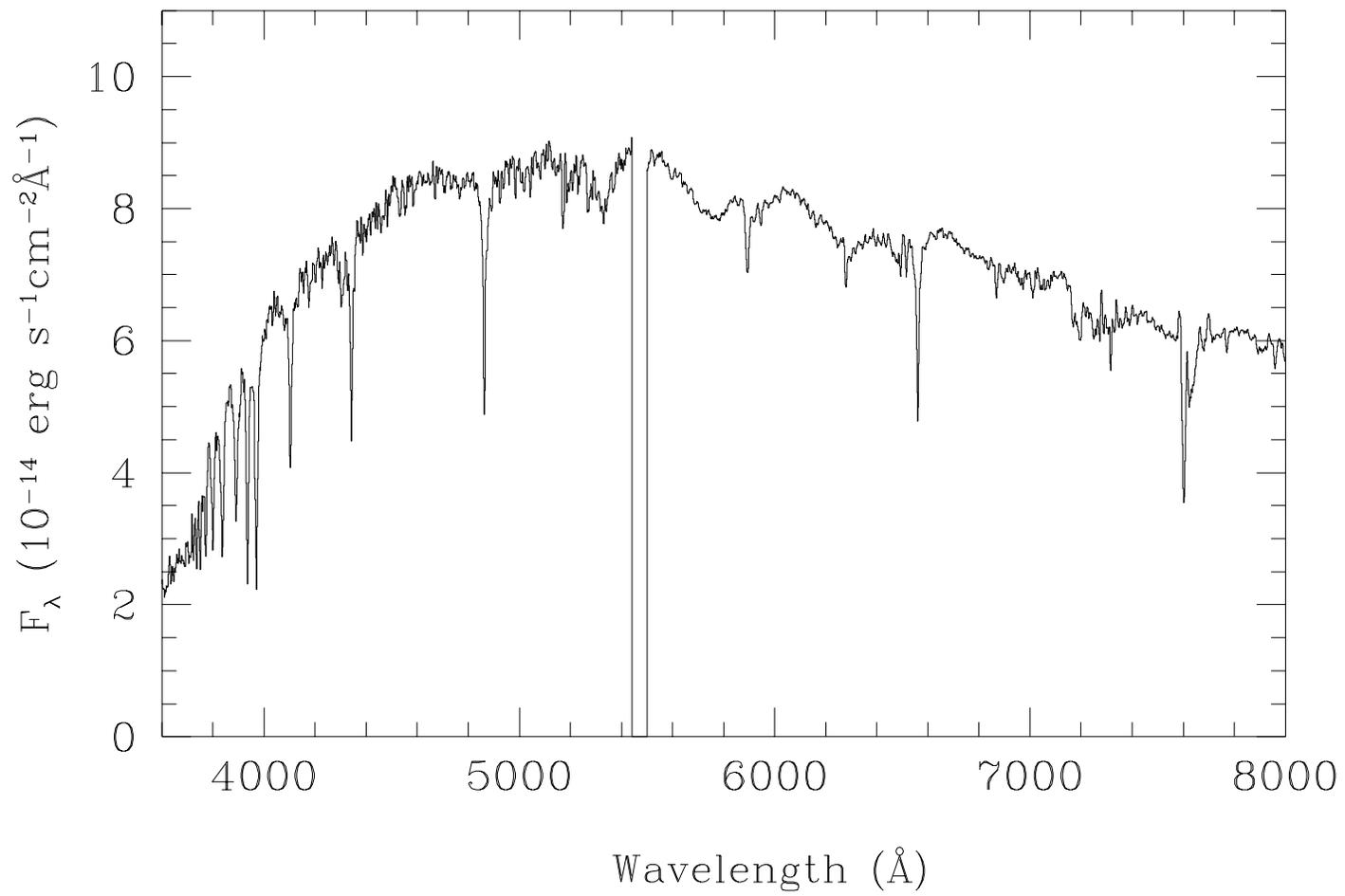
\end{document}